\def\gofr{g(r)}

\def\rone{{\bf r}_1}
\def\rtwo{{\bf r}_2}
\def\kbold{{\bf k}}
\def\rbold{{\bf r}}

\def\Gtwo{G_{cc}(r)}
\def\Stwo{S_{cc}(k)}
\def\vcoll{v_{coll}(r)}
\def\omegacoll{\omega_{coll}(k)}

\documentclass[aps,prb,preprint,superscriptaddress]{revtex4}
\usepackage{graphics}

\begin{document}

\markboth{Authors' Names}
{Paper's Title}

%
%

\title{The physics of liquid Para-Hydrogen\\}

\author{Thomas \ Lindenau, Manfred L.\ Ristig}
\address{Institut f\"ur Theoretische Physik, Universit\"at zu K\"oln\\
D--50937 K\"oln, Germany\\
ristig@thp.uni-koeln.de}
\author{Klaus A.\ Gernoth}
\address{School of Physics and Astronomy, The University of Manchester\\
Manchester, United Kingdom\\
k.a.gernoth@man.ac.uk}
\author{Javier \ Dawidowski}
\address{Consejo Nacional de Investigationes Cientificas y Tecnicas,\\
Centro Atomico Bariloche and Instituto Balseiro,\\
Comision Nacional de Energia Atomica,\\
Universidad Nacional de Cuyo, (8400) Bariloche, Argentina\\
javier@cab.cnea.gov.ar}

\author{Francisco J.\ Bermejo}
\address{C.S.I.C. Dept. of Electricity and Electronics,\\
University of the Basque Country, P.O. 644, E-48080 Bilbao, Spain\\
javier@langran.iem.csic.es}


\begin{abstract}
  Macroscopic systems of hydrogen molecules exhibit a rich
  thermodynamic phase behavior.  Due to the simplicity of the
  molecular constituents a detailed exploration of the thermal
  properties of these boson systems at low temperatures is of
  fundamental interest. Here,we report theoretical and experimental
  results on various spatial correlation functions and corresponding
  distributions in momentum space of liquid para-hydrogen close to the
  triple point. They characterize the structure of the correlated
  liquid and provide information on quantum effects present in this
  Bose fluid. Numerical calculations employ Correlated
  Density-Matrix(CDM)theory and Path-Integral
  Monte-Carlo(PIMC)simulations. A comparison of these theoretical
  results demonstrates the accuracy of CDM theory. This algorithm
  therefore permits a fast and efficient quantitative analysis of the
  normal phase of liquid para-hydrogen.We compare and discuss the
  theoretical results with available experimental data.
\end{abstract}


\maketitle

\section{Introduction}

Microscopic, mesoscopic, and macroscopic systems of hydrogen atoms or
molecules are of fundamental importance in quantum many-body theory.
The properties of a single hydrogen atom led the basis for the atomic
shell model. A single hydrogen molecule is the simplest compound and
has been an early laboratory for studying chemical binding. Its
spectrum exhibits the effects of internal degrees of freedom leading
to different levels of excitations and the existence of ortho- and
para-hydrogen. The molecule therefore shares common features with
nuclei where internal degrees are important and cause violation of the
so-called Y-symmetry\cite{1,2}. Atomic hydrogen gas in a strong
magnetic field remains a Bose fluid even at zero temperature and can
condense into a Bose-Einstein phase under certain conditions\cite{3}
like the alkali gases\cite{4}.

Liquid and solid phases of molecular hydrogen and deuterium have been
extensively studied, experimentally as well as theoretically. These
many-body systems are of continuing interest for a variety of reasons.
Hydrogen molecules are the dominant constituents of giant
planets\cite{5}. The physics of metallic hydrogen is explored by many
researchers to achieve the insulator-metal cross-over\cite{6}.

The hydrogen liquid, close to the triple point, is the object of
current research. In this paper we concentrate on a theoretical and
experimental analysis of its quantum properties close to the triple
point. There are many other open and interesting questions, such as
properties of mesoscopic clusters and films, molecular hydrogen in
confined geometries, Bose- Einstein condensation in solid and
supercooled liquid hydrogen, etc.

We investigate the properties of the one- and two-body reduced
density-matrix elements of liquid para-hydrogen in its normal boson
phase at low temperatures.The associated Fourier transverse of these
quantities reveal the spatial structure of the correlated system.
Detailed numerical calculations are performed at the temperature $T =
16\,{\rm K}$ and a particle number density $\rho=0.021\,$\AA$^{-3}$. The
experimental measurements for the liquid structure function\cite{7}
and the dispersion law of the collective excitations\cite{8} have been
done at the temperature $T = 15.2\,{\rm K}$. Quantitative information
on single-particle properties have been gained by precision
neutron-scattering experiments at $T=15.7\,{\rm K}$ and various
pressures that permits to extract the momentum distribution of a
molecule in the hydrogen liquid\cite{9}.

The present theoretical analysis is based on the parameter-free
microscopic CDM theory\cite{10,11,12} and PIMC calculations\cite{13}
with the central Silvera-Goldman potential as input\cite{14}. Section
2 begins with a quantitative study of the spatial distribution
function $g(r)$ and the static structure function $S(k)$ comparing the
CDM results with the corresponding PIMC results. Theoretical results
on the excitation energies and quasiparticle energies within CDM
theory are discussed and compared with the measured dispersion laws in
Section 3. The theoretical momentum distribution $n(k)$ is analysed in
Section 4. Its study is based on a structural factorization of the
one-body reduced density-matrix elements $n(r)$ reported
earlier\cite{15} that permits a clean separation of particle exchange
properties from spatial phase-phase correlations caused by the
intermolecular interactions. A short summary is given in Section 5.

\section {Spatial Structure}

The spatial structure of a homogeneous quantum fluid in thermal
equilibrium may be characterized by a set of correlation functions and
concomitant Fourier transforms or structure functions.The radial
distribution function $g(r)$ describes the spatial correlations
between two particles in coordinate space and depends on the relative
distance between them, $r=\left|\rone-\rtwo\right|$.The associated
static structure function $S(k)$ is the dimensionless Fourier
transform of $g(r)$.Its dependence on the relative momentum or
wavenumber k can be extracted from measured neutron scattering cross
sections.The quantum-mechanical correlations between identical
particles become apparent in the cyclic or particle-exchange
correlation function $\Gtwo$ and its Fourier inverse $\Stwo$.
Information on the spatial correlations between the phase factors
associated with the single-particle wave components of the full N-body
density matrix of the thermodynamic state of the quantum fluid is
embodied in the phase-phase correlation function $Q(r)$. This quantity
is contained in the off-diagonal elements $n(r)$, i.e., the one-body
reduced density matrix.

\begin{figure}[th]
\begin{centering}
\resizebox{1\textwidth}{!}{\includegraphics{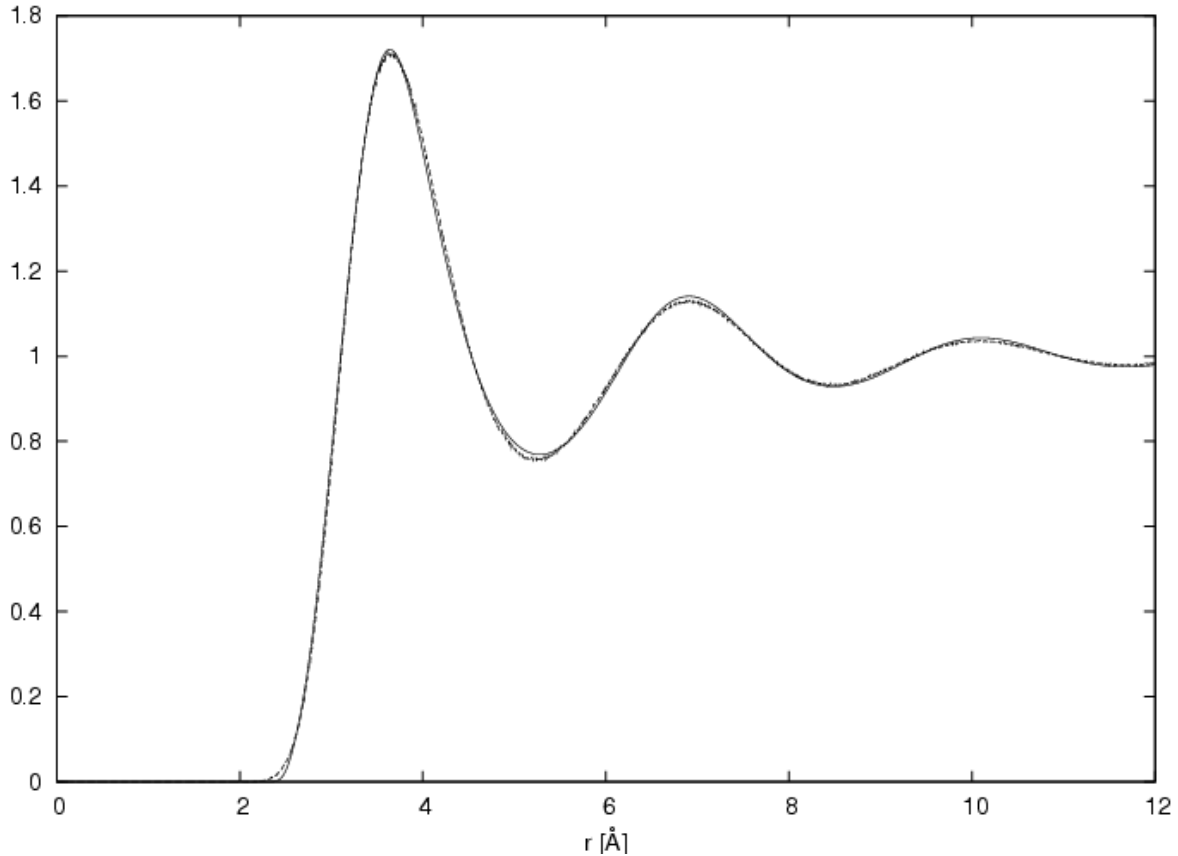}}
\caption{Radial distribution function $g(r)$ of liquid para-hydrogen
  ($T=16$ K, $\rho =0.021$ \AA$^{-3}$).  Full line: results of CDM theory,
  broken line: PIMC simulation results.}
\end{centering}
\end{figure}

This Section deals with a thorough analysis and comparison of
theoretical and experimental results on functions $g(r)$ and $S(k)$.
CDM theory provides a renormalized Schr\"odinger equation with zero
eigenvalue for the square root of the radial distribution function,
\begin{equation}
  \left[-{\hbar^2\over m}\Delta + v(r) + w(r) + v_{coll}(r) + v_{qp}(r)\right]\sqrt{\gofr} = 0\, .
\label{E1}
\end{equation}
The potential energy terms appearing in Eq.~(\ref{E1}) are the Silvera
potential\cite{14} $v(r)$, the induced potential\cite{12} $w(r)$, the
coupling term $\vcoll$ to the collective excitations\cite{10}, and the
quasiparticle coupling term\cite{12} $v_{qp}(r)$. For liquid
para-hydrogen close to the triple point the last coupling term is very
small and may be ignored.  Without the term $v_{qp}(r)$ it is
straight-forward to solve the Schr\"odinger equation by following the
elementary calculus designed in Ref.~\onlinecite{10}. The solution is
displayed in Figure 1. We have checked its numerical accuracy by
comparing the CDM results with those of a PIMC calculation for the
same temperature and density\cite{16}, finding excellent agreement(see
Figure 1).

\begin{figure}[b]
\resizebox{1\textwidth}{!}{\includegraphics{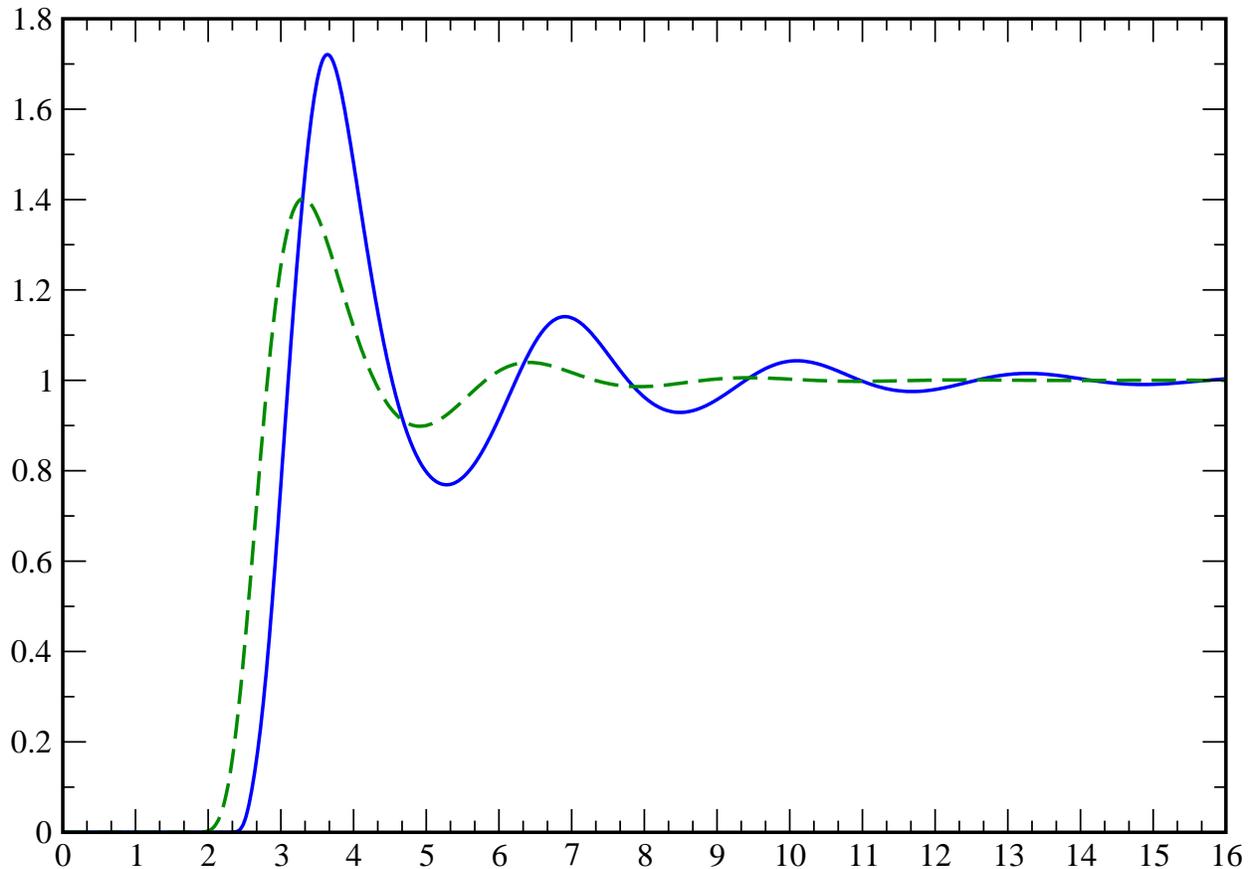}}
\vspace{1cm}
\caption{Comparison of radial distribution functions of liquid
  para-hydrogen and of normal liquid helium (broken line) under
  similar thermodynamic conditions (abscissa: relative distance $r$ in
  units [\AA]). }
\end{figure}

The results on the distribution $\gofr$ show the enormous correlation
strength that exists in the hydrogen liquid. The first maximum at
$r=3.7$\,\AA\ attains a value that is significantly larger than the
maximal strength one finds at the first peak of the radial
distribution function of liquid helium under similar thermal
conditions assuming the realistic Aziz potential\cite{17} as input for
the theoretical analysis (Figure~2).

Next, we turn to a detailed study of the associated static structure
function $S(k)$. Within CDM theory as well as in the stochastic PIMC
approach this quantity is calculated as the dimensionless Fourier
transform

\begin{equation}
  S(k) = 1 + \rho\int\left[\gofr -1\right] {\rm e}^{i\kbold\rbold} d\rbold\, .
\label{E2}
\end{equation}

Numerical results on the structure function with CDM and PIMC data for
$g(r)$ as input are displayed in Figure 3.  Once more,the Monte-Carlo
results confirm the high accuracy of the results from CDM theory for
liquid para-hydrogen close to the triple point. We emphasize that the
integral (\ref{E2}) has been calculated without applying any fitting
prescription for the distribution $g(r)$ at large relative distances.
The oscillations seen in the stochastic results at wavenumbers $k <
1.5$\,\AA$^{-1}$ are artifacts of the inherently unavoidable small box
size used in the PIMC calculations.  In contrast, CDM theory does not
suffer from this deficiency since we can easily solve the Schr\"odinger
equation in a sufficiently large interval of relative distances. CDM
theory provides therefore an efficient tool for accurate calculations
of the isothermal compressibility and isothermal velocity of sound in
the limit of vanishing wavenumber.

\begin{figure}[h]
\resizebox{1\textwidth}{!}{\includegraphics{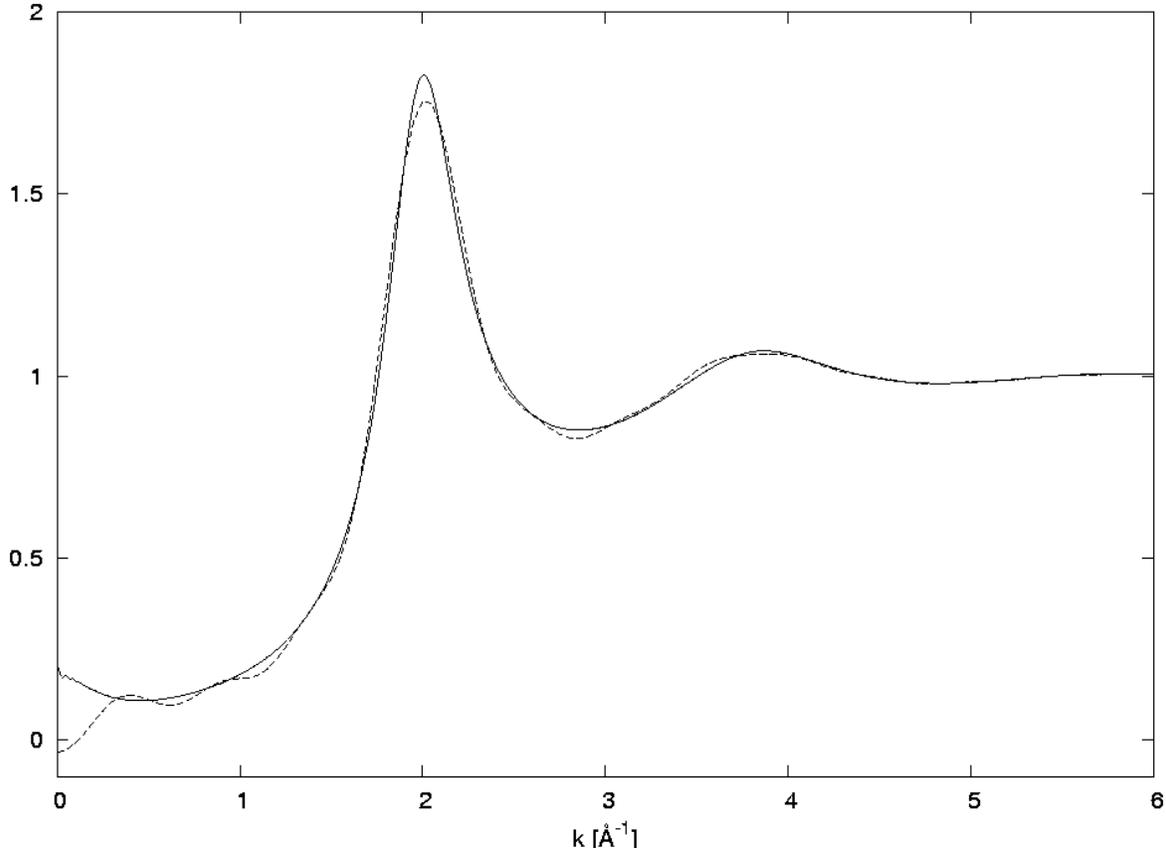}}
\caption{ Numerical results on the structure function $S(k)$ of liquid
  para-hydrogen by CDM theory and from PIMC calculations (dashed
  line), see text.}
\end{figure}

The theoretical results reported may be compared with experimental
data from neutron scattering experiments\cite{7}.  The data processing
yields an experimental structure function with a much higher first
peak than found in the theoretical result on $S(k)$. This discrepancy
between peak values of about 2.8 experimentally and about 1.8 for the
theoretical result is not acceptable. There are theoretical data from
molecular dynamics calculations\cite{7} which seem to support the
experimental findings (see Figures 3 and 5 of Ref.~\onlinecite{7}).
However, they are evidently in disagreement with the present results
within CDM theory and of PIMC calculations. This problem -at least on
the theoretical side- awaits a future careful examination.

\section{Excitations}

A correlated normal Bose fluid permits at least two different branches
of excitations.There is a collective branch of phonons at small
wavenumbers k and possibly roton-like excitations at atomic
wavelengths. A second branch consists of excited quasiparticles that
conserves the total number of constituents.CDM theory enables us to
determine the excitation energies of both branches by solving two
Euler-Lagrange equations\cite{12},
\begin{equation}
{\delta F_\lambda\over\delta n_{cc}(k)}=0\, ,
\label{E3}
\end{equation}
with subcondition
\begin{equation}
\sum_\kbold n_{cc}(k) = N
\label{E4}
\end{equation}
and
\begin{equation}
{\delta F_\lambda\over\delta n_{coll}(k)}=0\, .
\label{E5}
\end{equation}
The thermodynamic potential $F_\lambda [S(k),n_{cc}(k),n_{coll}(k);T,\rho ]$
is a functional of the static structure function $S(k)$, the momentum
distribution $n_{cc}(k)$ of $N$ quasiparticles, and the
occupation-number density $n_{coll}(k)$ of the collective excitations.
An explicit expression for the functional $F_\lambda$ can be constructed
within CDM theory by following a systematic approximation procedure.
At the present level of formal development the functional is generated
from a trial $N$-body density matrix of Jastrow- type. Future
improvements may be derived by a suitable generalization of the
Correlated Basis Functions (CBF) formalism for the ground state of a
quantum many-body system\cite{18,19}.

We have solved the Euler-Lagrange equations using the approximate
thermodynamic potential derived from the Jastrow trial $N$-body
density matrix\cite{12}. At the assumed temperature $T=16$ K and
particle-number density $\rho=0.021$ \AA$^{-3}$ one finds that the optimal
momentum distribution of the quasiparticles is excellently represented
by the Gaussian form
\begin{equation}
n_{cc}(k) = \exp{\beta\left[\mu_0 - \epsilon_0(k)\right]}
\label{E6}
\end{equation}
with $\epsilon_0(k) = \hbar^2 /2m$ (molecular mass m of hydrogen) and inverse
temperature $\beta = (k_B T)^{-1}$.  The chemical potential is $\mu\simeq
-7.8$ K due to the subcondition (\ref{E4}). The coresponding one-body
reduced density matrix is therefore given by
\begin{equation}
n_{cc}(r) = \exp{\left[ -\pi \left({r\over\lambda}\right)^2\right]}
\label{E7}
\end{equation}
with a thermal wavelength $\lambda\simeq 3.09$ \AA. A consequence of this result
is the absence of particle-exchange correlations in the hydrogen state
considered, explicitly expressed by the result\cite{11,12} $\Stwo =
\Gtwo\simeq 0$. The quasiparticles are therefore distinguishable free
hydrogen molecules obeying classical Boltzmann statistics.

\begin{figure}[h]
\resizebox{1\textwidth}{!}{\includegraphics{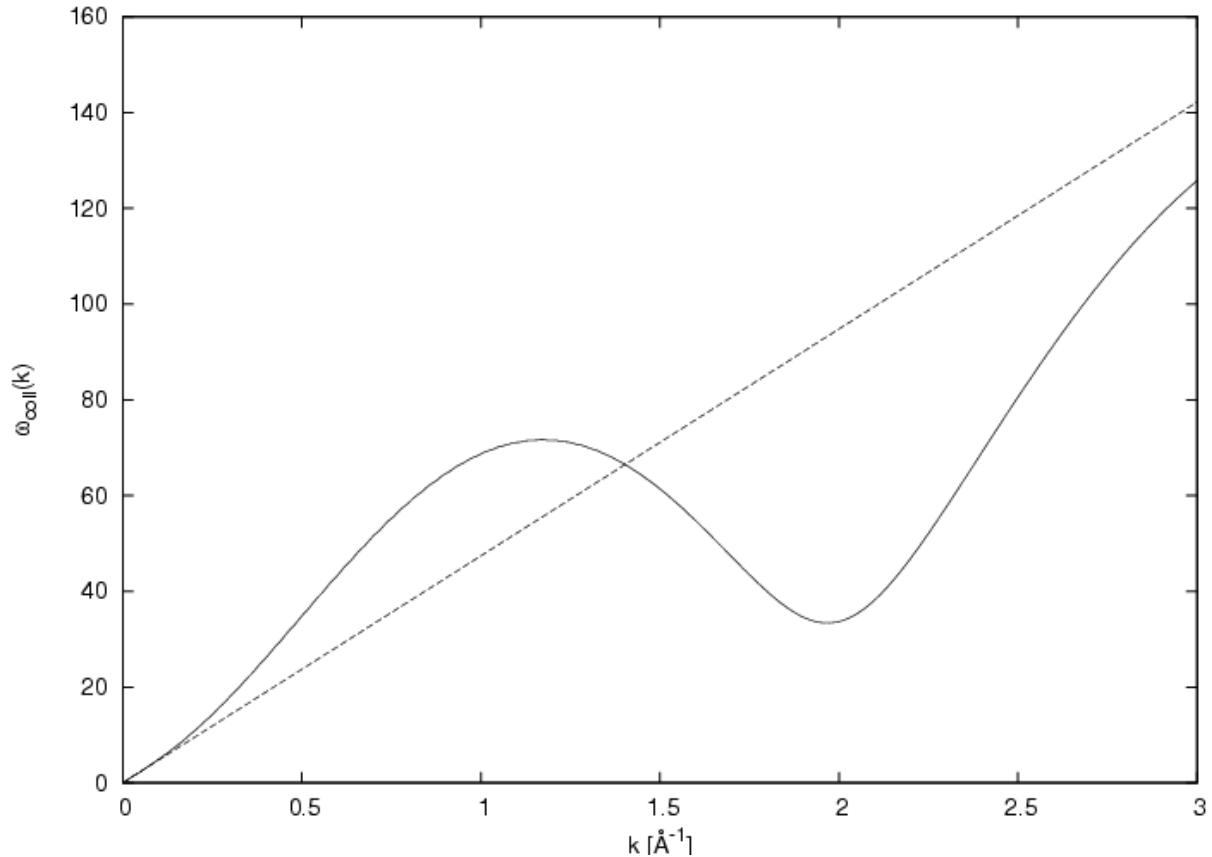}}
\caption{ Energy spectrum of collective excitations in the hydrogen
  system, in generalized Feynman approximation.The dashed line
  indicates the linear phonon spectrum in the limit of vanishing
  wavenumber.}
\end{figure}

Eq.~(\ref{E5}) is a generalized Feynman equation\cite{12,20} that can
be easily solved. The optimal excitation energies $\omegacoll$ are
displayed in Figure 4. This energy branch is of the familiar
phonon/roton form. The broken line indicates the phonon limit
$\omegacoll\simeq\hbar ck$ as $k$ approaches zero.  Its slope yields the
theoretical isothermal sound velocity $c \simeq 680$ m/s. We may compare
these results with the available experimental data \cite{8} (cf.
Figures 5 and 8 therein).  There is good agreement between the
theoretical and the experimental results in the roton region. However,
the roton excitations are strongly damped.In contrast,the excitations
in the wavenumber region $k < 0.8$ \AA$^{-1}$ are stable, yet the
theoretical excitation energies become significantly smaller than the
experimental energies by increasing the wavenumber $k$. The
discrepancy requires further investigation.

\section{Momentum Distribution}
 
The momentum distribution of a single hydrogen molecule in the liquid
is given by the integral

\begin{equation}
n(k) = \rho\int n(r) {\rm e}^{i\kbold\rbold} d\rbold
\label{E8}
\end{equation}
where function $n(r)$ is the (unit-normalized) one-body reduced
density matrix. This quantity has been analyzed within CDM
theory\cite{2,15}. The formalism yields the structural factor
decomposition
\begin{equation}
n(r) = n_c N_0(r) \exp\left[ -Q(r) \right]
\label{E9}
\end{equation}
with the strength factor $n_c = \exp Q(0)$. The factor $N_0(r) =
n_{cc}(r) + N_c(r)$ embodies the quantum-mechanical effects due to the
exchange of identical particles. The function $P(r)= Q(r)/Q(0)$ can be
interpreted as the (unit-normalized) phase-phase correlation function.

\begin{figure}[ht]
\resizebox{1\textwidth}{!}{\includegraphics{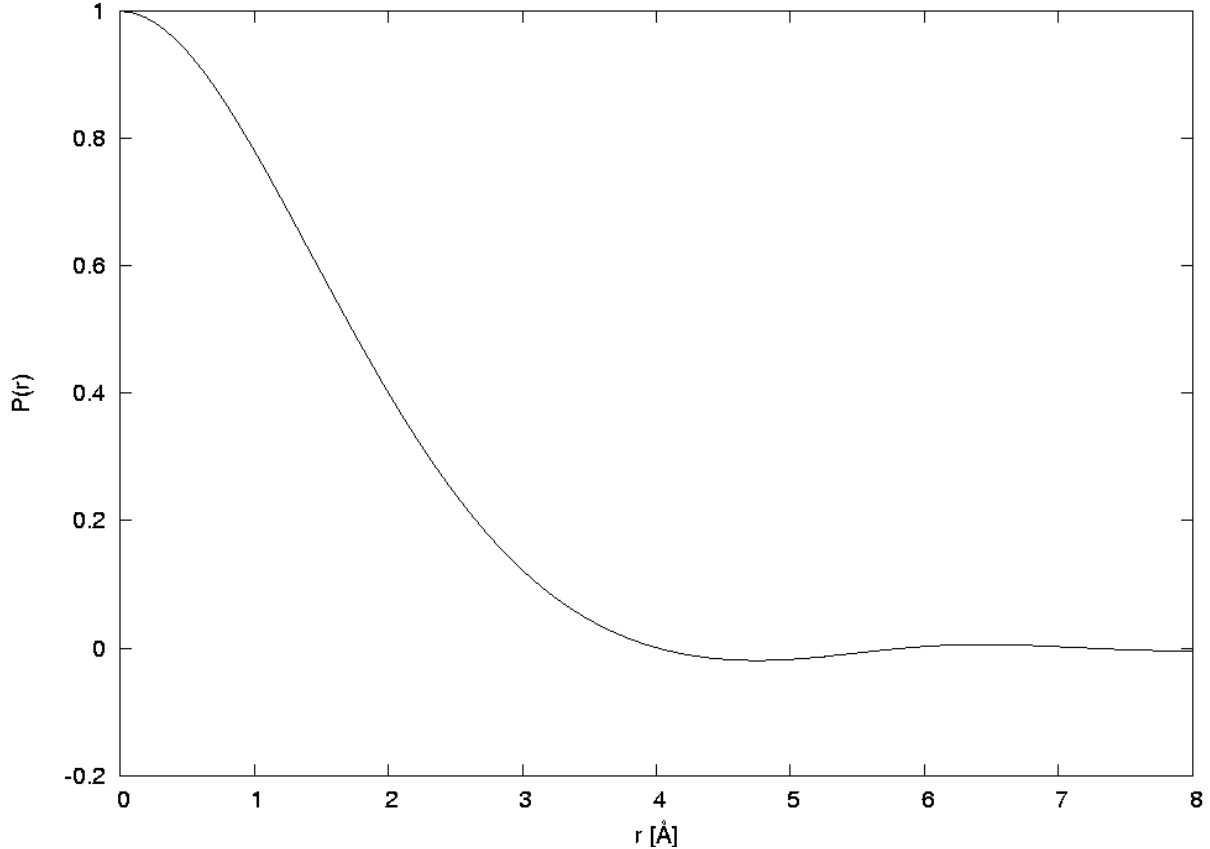}}
\caption{ Numerical results for the phase-phase correlation function
  $P(r)$ by CDM theory, in HNC/0 approximation.}
\end{figure}

Employing the hypernetted-chain (HNC) technique for evaluating the
functions\cite{15} we calculated $N_0(r)$ and $P(r)$ for liquid
hydrogen in HNC/0 approximation. As expected, the exchange
correlations $N_c(r)$ are very small and thus
\begin{equation} 
N_0(r) \simeq n_{cc}(r)\, .
\label{E10}
\end{equation}
Figure 5 displays theoretical results on function {P(r)} in HNC/0
approximation.  We see that this distribution is well represented by a
Gaussian form
\begin{equation}
P(r)\simeq P_G(r) = \exp{\left[ -\pi \left( {r\over\lambda_P }\right)^2 \right]}\, ,
\label{E11}
\end{equation}
with $\lambda_P\simeq 3.73$\,\AA\ at $T=16$ K and $\rho = 0.021$\,\AA$^{-3}$.  The
strength factor $n_c$ is related to the curvature of function $P(r)$
and the total kinetic energy per molecule of the Bose liquid.  A PIMC
calculation of this energy portion yields 58.6 K at $T=16$ K and $\rho =
0.021$\,\AA$^{-3}$.  With these input data the strength factor has the
value $n_c = 0.118$.

\begin{figure}[h]
\resizebox{1\textwidth}{!}{\includegraphics{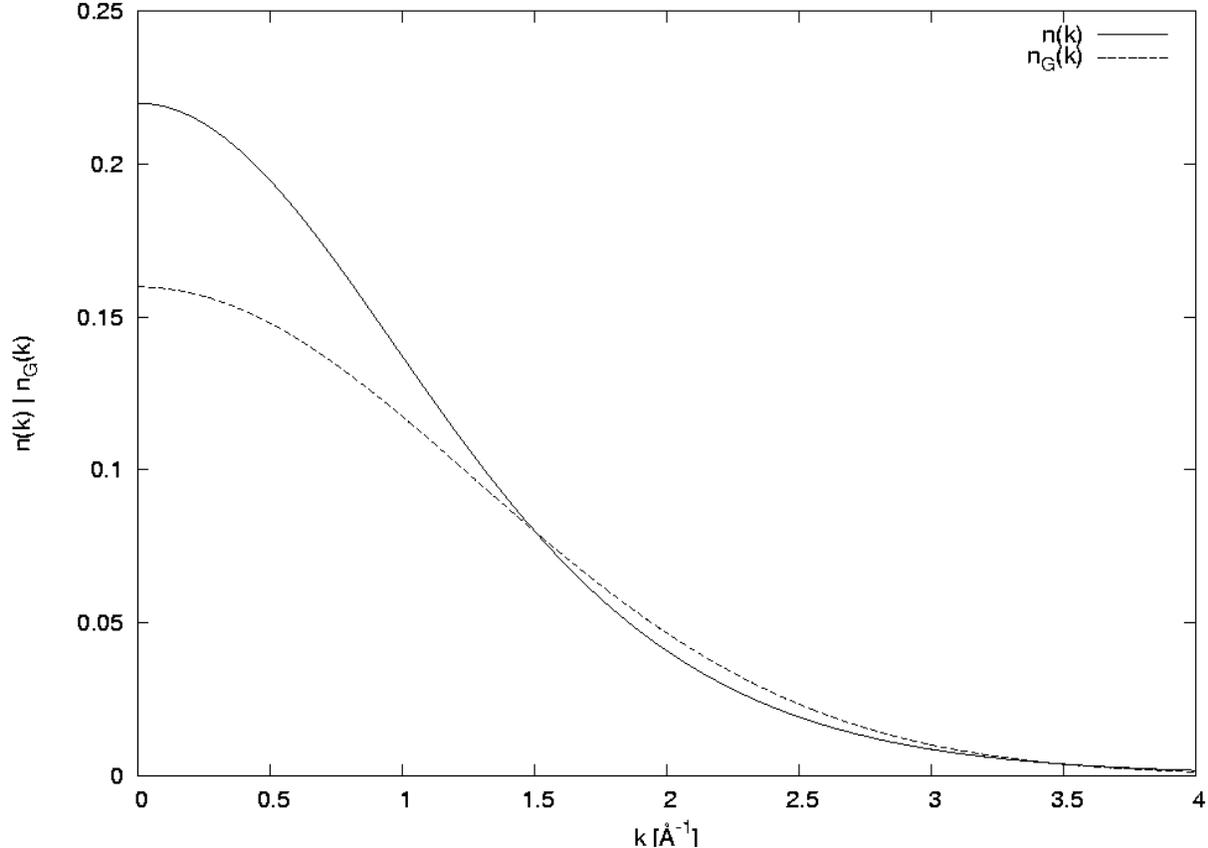}}
\vspace*{8pt}
\caption{ Theoretical momentum distribution $n(k)$ of a single
  hydrogen molecule in the liquid. It is compared to the classical
  Gauss distribution $n_G(k)$ that leads to the same total kinetic
  energy per molecule as the correlated hydrogen liquid ( broken
  line).}
\end{figure}

Straightforward multiplication and integration via Eq.~(\ref{E9})
yield the one-body elements $n(r)$ and the momentum distribution
$n(k)$. Figure 6 represents numerical results on the dependence of the
latter function on wavenumber $k$.  We may compare it with the
Gaussian momentum distribution that would hold by assuming that the
molecules are free and distinguishable but have the same total kinetic
energy per particle as the quantum liquid. This distribution would be
characterized by an effective thermal wavelength $\lambda_0\simeq 1.97$\,\AA\
(Figure 6, broken line). Obviously, the difference
\begin{equation}
\Delta N(k) = k^2\left[ n(k) - n_G(k) \right]
\label{E12}
\end{equation}
measures the deviation of the kinetic energy distribution of the
quantum liquid from the classical Maxwell distribution of
distinguishable hydrogen molecules. Its dependence on wavenumber $k$
is displayed in Figure 7. We see that the quantum-mechanical energy
distribution of the system favors an increase of the number of
molecules with low momenta compared to the classical distribution.

\begin{figure}[t]
\resizebox{1\textwidth}{!}{\includegraphics{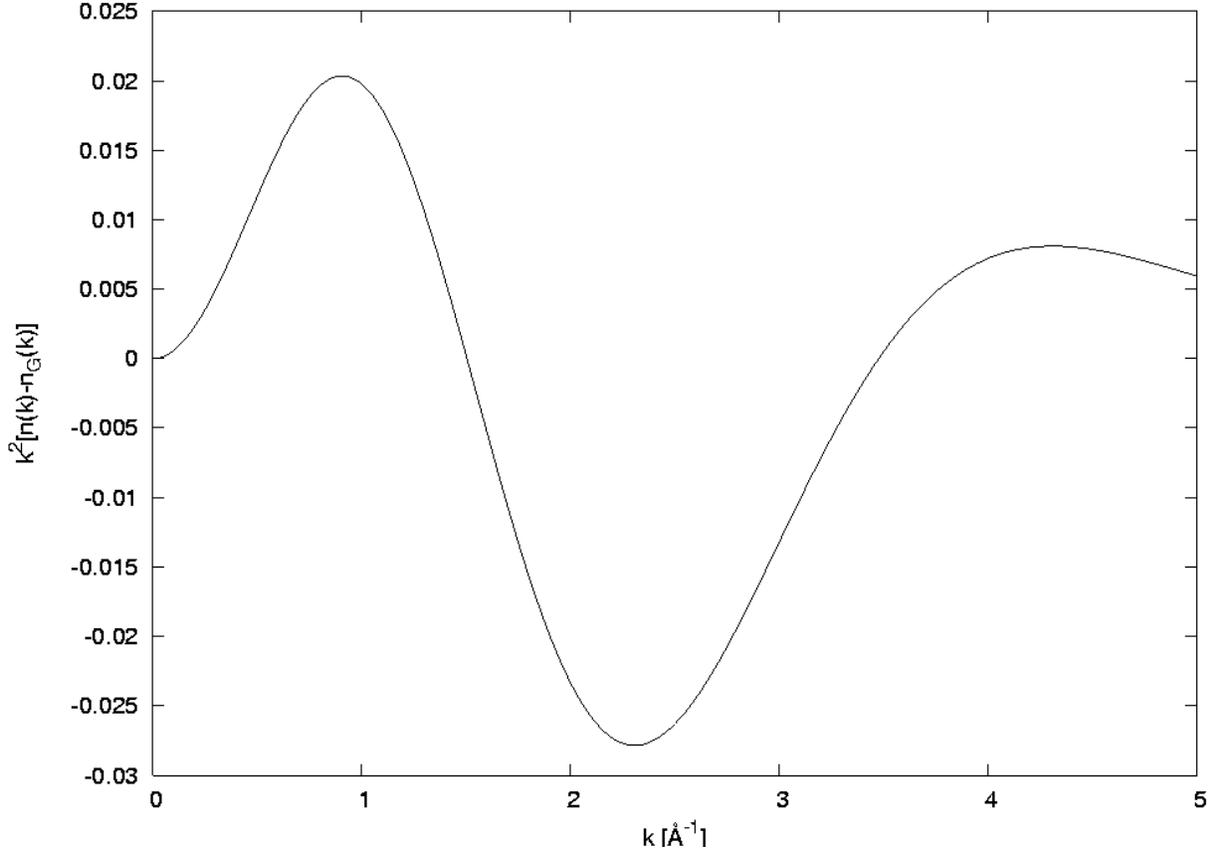}}

\caption{ Deviation of the kinetic energy distribution $k^2n(k)$ of
  liquid para-hydrogen from the corresponding classical Maxwell energy
  distribution (cf.~Eq.~(\ref{E12})).}
\end{figure}

\section{Summary} 

We have presented a theoretical and an experimental analysis of the
structure of liquid para-hydrogen close to the triple point.The system
is characterized by very strong dynamical spatial correlations induced
by the intermolecular forces. However, the repulsion at short relative
distances suppresses particle-exchange correlations between and among
the hydrogen molecules. For the same reason, the exchange (or cyclic)
correlation function and the associated exchange structure function
are almost zero everywhere. Furthermore, the quasiparticle momentum
distribution is very well approximated by the classical Gaussian
distribution of free and distinguishable particles. Similarly,the
short-ranged phase-phase correlation function and corresponding
structure function are to a very good approximation of Gaussian form.
The interplay of exchange and phase-phase coupling, however, generates
a significant departure of the single-particle momentum distribution
in liquid para-hydrogen from the classical Maxwell-Gauss distribution.

The present study employed the formalism of CDM theory for a normal
Bose fluid. The results have been compared with PIMC simulation data.
The numerical comparison demonstrates the high accuracy of CDM theory
when applied to liquid hydrogen at low temperatures. CDM theory is
therefore expected to be a very fast, efficient, and reliable tool for
a detailed quantitative analysis of normal Bose fluids under similar
thermodynamic conditions.

\end{document}